\documentclass[prl,aps,showpacs,tightenlines,nofootinbib]{revtex4}
\usepackage{graphicx}
\usepackage{amsfonts}
\usepackage{amssymb}
\usepackage{latexsym}

\newcommand{\CL}{{\cal L}}

\newcommand{\CG}{{\cal G}}

\newcommand{\bear}{\begin{array}}  \newcommand{\eear}{\end{array}}
\newcommand{\bea}{\begin{eqnarray}}  \newcommand{\eea}{\end{eqnarray}}
\newcommand{\beq}{\begin{equation}}  \newcommand{\eeq}{\end{equation}}
\newcommand{\bef}{\begin{figure}}  \newcommand{\eef}{\end{figure}}
\newcommand{\bec}{\begin{center}}  \newcommand{\eec}{\end{center}}
\newcommand{\non}{\nonumber}  
\newcommand{\lmk}{\left(}  \newcommand{\rmk}{\right)}
\newcommand{\lkk}{\left[}  \newcommand{\rkk}{\right]}
  
\newcommand{\lnk}{\left \{ }  \newcommand{\rnk}{\right \} }
  
\newcommand{\del}{\partial}  
\newcommand{\vect}[1]{\mbox{\boldmath${#1}$}}
\newcommand{\vecs}[1]{\mbox{\boldmath\tiny${#1}$}}
\newcommand{\bib}{\bibitem} \newcommand{\new}{\newblock}
\newcommand{\la}{\left\langle} \newcommand{\ra}{\right\rangle}

\newcommand{\mg}{M_G}

\newcommand{\phip}{\phi_{+}}
\newcommand{\phim}{\phi_{-}}
\newcommand{\phic}{\phi_c}

\newcommand{\im}{\mbox{Im}}
\newcommand{\re}{\mbox{Re}}

\newcommand{\vek}{\vect k}
\newcommand{\vep}{\vect p}

\newcommand{\op}{\omega_p'}

\def\PRD#1#2#3{Phys. Rev. D {\bf #1}, #2 (19#3)}

\begin{document}
\title{Thermal background can solve the cosmological moduli problem}

\author{Jun'ichi Yokoyama} 
\affiliation{Research Center for the Early Universe (RESCEU), Graduate
School of
Science, The University of Tokyo, Tokyo 113-0033, Japan}

\date{\today}


\begin{abstract}
It is shown that the coherent field oscillation of moduli fields
with weak or TeV scale masses
can dissipate its energy efficiently if they have a derivative
coupling to standard bosonic fields in a thermal state.
This mechanism may provide a new solution to the cosmological moduli
problem in some special situations \cite{go}.
\end{abstract}


\maketitle

\tighten

Modern theories of high energy physics contains a number of 
scalar fields which have a flat potential and
interacts with ordinary particles only with
the gravitational strength \cite{ellis}.
In the context of low-energy supersymmetry 
these moduli fields
 typically have a flat potential intrinsically and acquire a
mass of order of weak or TeV scale when supersymmetry is broken 
\cite{deCarlos:1993jw}, although other mechanisms of
 moduli stabilization have also been 
extensively discussed recently \cite{Giddings:2001yu}.


During inflation in the early universe
\cite{Sato:1980yn}, supersymmetry
is spontaneously broken in a different manner than it is today
due to the large vacuum energy density.  Then moduli fields,
which we denote by $\phi$, 
typically acquire
a mass of order of the Hubble parameter and they relax to
 a potential minimum at this stage,
which is deviated from today's  value at the current potential minimum by 
up to the gravitational scale, 
$\Delta\phi \lesssim \mg=2.4\times 10^{18}$ GeV.
After inflation their mass is turned off to a much smaller value
due to the disappearance
of vacuum energy density and they do not move until the 
Hubble parameter decreases to their eventual mass scale which 
is presumably of order of the weak scale or TeV scale as stated above.
The scalar fields then start coherent oscillation with the 
initial amplitude up to $\mg$, which 
will dominate the energy density
of the universe eventually.  
According to the conventional estimates,  
their lifetime is given by
\beq
  \tau_\phi \approx \frac{\mg^2}{m_\phi^3} \approx
 10^8\lmk\frac{m_\phi}{10^2\rm GeV}\rmk^{-3}{\rm sec}, \label{decayold}
\eeq
that is, they decay after the primordial nucleosynthesis
creating a huge amount of entropy to demolish the successful
primordial nucleosynthesis \cite{ellis}.


In this Letter we present a new class of solution to the above
cosmological moduli problem by 
arguing that the previous estimate of the decay 
rate (\ref{decayold}), which has been used in all the other 
proposed solutions
to the problem \cite{old}, does not apply in a finite-temperature and
finite-density state of the early universe and that it is much 
more enhanced than in the case of decay in a vacuum.
 As a result
we show that the coherent moduli oscillation may efficiently 
dissipate its energy density well before the big bang nucleosynthesis
in some special situations.

The crucial point is to take moduli decay through
 derivative coupling correctly into
account, such as a coupling with a kinetic term of other fields.
Indeed we expect moduli field is coupled with gauge fields through
$\frac{\phi}{\mg} F_{\mu\nu}F^{\mu\nu}$.  It may also be coupled with 
scalar fields as $\lambda\frac{\phi}{\mg} (\partial \chi)^2$ or 
$\lambda\frac{\phi}{\mg} \chi\Box \chi$ where $\chi$ is a generic 
scalar field and $\lambda$ is a dimensionless coupling constant
of order unity \cite{Moroi:1999zb}.
Previously, these couplings were expected to give a decay rate no
different from (\ref{decayold}) at most, because using the
equation of motion,
$
  \Box \chi =m_\chi^2\chi, 
$
it was concluded that the derivative coupling would give
a decay rate similar to the coupling $\frac{\phi}{\mg}m_\chi^2\chi^2$,
which yields (\ref{decayold}) for $m_\phi \gtrsim 2m_\chi$.

Such a naive analysis, however, could only be valid in decay in 
the vacuum and would not apply in the high-temperature environment
in the early universe.  If $\chi$ is a standard field in the visible
sector, it is 
strongly coupled with other degrees of freedom in the early 
universe and rapidly reaches thermal equilibrium.  Then they 
acquire a thermal mass of order of $\sim gT$ in general, where
$g$ is a typical gauge coupling and $T$ is the cosmic temperature,
so that the right-hand-side of the above equation of motion
 could be significantly 
enhanced.
Alternatively, one may regard the derivative $\partial$ acting on
$\chi$ as not yielding its rest mass energy $m_\chi$ but 
energy-momentum arising from finite-temperature environment, 
$\partial \sim T$ modulo some coupling.  Then,
the strength of 
interaction $\lambda\frac{\phi}{\mg}(\partial \chi)^2$
is estimated as
\beq
  {L}_{\rm int} \approx \frac{\lambda(gT)^2}{\mg}\phi\chi^2.
\label{effectiveint} 
\eeq
The decay rate of $\phi$ through the above interaction
should read
\beq
 \Gamma \approx \frac{\lambda^2g^4T^4}{8\pi\mg^2m_\phi}
 \lkk 1+ 2n_B\lmk \frac{m_\phi}{2}\rmk\rkk C 
 \approx \frac{\lambda^2g^4T^5}{2\pi\mg^2m_\phi^2}C,
\label{rough}
\eeq
where $C$ is a suppression factor due to a large thermal
mass of the decay product $\chi$ which has been given 
in \cite{Yokoyama:2005dv}
for a specific model.  Here $n_B(\omega)=1/(e^{\omega/T}-1)$ is
the thermal number density of a boson and the factor in the
bracket represents the effect of induced emission
\cite{Boyanovsky:1995em,Yokoyama:2004pf}.
Taking $\lambda\sim g \sim 1$, $m_\phi\sim 10^2$ GeV, 
$T\sim 10^{10}$ GeV, which is a typical radiation temperature
at the onset of moduli oscillation $H\sim m_\phi$,
we find $\Gamma \sim 3\times10^8C$ GeV.  Thus if $C$ takes an appropriate
value, $\phi$ can dissipate its energy right after it starts
oscillation.

In order to examine if the above naive estimate is correct, we employ 
nonequilibrium field theory to calculate the dissipation rate
of a modulus $\phi$ in the presence of a derivative interaction.
For simplicity we consider the following model consisting of
two scalar fields, $\phi$ and $\chi$.
\beq
  {\cal L}=\frac12\,(\del_{\mu}\phi)^2-\frac12\,m_{\phi}^2\phi^2 
        + \frac12\,(\del_{\mu}\chi)^2-\frac12\,m_{\chi}^2\chi^2 
    +\lambda\frac{\phi}{\mg}(\partial\chi)^2 - \CL_{int}
      \:,  \label{lagrangian}
\eeq
where $\CL_{int}$ is interaction of $\chi$ which thermalizes it.

We calculate the dissipation rate of $\phi$ under the following
setup appropriate to the specific problem we are working on.
First we neglect
cosmic expansion since we are interested only
in the case dissipation time is
shorter  than the cosmic expansion time.  Second we 
assume $\chi$ is in a thermal state with a specific temperature
$T=\beta^{-1}$ owing to the rapid thermalizing interaction due
to the self coupling.  Finally, 
 we consider the situation the parametric resonance
 \cite{Traschen:1990sw} is 
ineffective which is the case for the modulus mass
range of our interest \cite{Shuhmaher:2005mf}.

We 
calculate an effective action for $\phi$ and derive an equation of
motion for its expectation value using the closed time-path
formalism \cite{Sch,Chou:es}. 
Although this method has been applied to various cosmological 
problems by a
number of authors \cite{Mor,mm}, to our knowledge, derivative
coupling at finite temperature has not been investigated in this context
yet.
The one-loop effective action relevant to dissipation due to the
derivative coupling is given by
\bea
&&  \Gamma[\phi_{c},\phi_{\Delta}] =-\int d^4x 
        \phi_{\Delta}(x)( \,\Box+m_\phi^2\,)  \phi_{c}(x)
                                           \non \\
       && -\int d^4x d^4x' 
            C(x-x')
 \theta(t-t')\phi_{\Delta}(x)\phi_{c}(x') 
   +\frac{i}{2}\int d^4x d^4x' 
                D(x-x')
\phi_{\Delta}(x)\phi_{\Delta}(x')+\cdots
 \:.
  \label{effectiveaction}
\eea
Here $\phic$ and $\phi_{\Delta}$ are mean and difference of the
field variable in the forward time branch ($t=-\infty$ to
$+\infty$), $\phip$, and that in the backward time branch ($t=+\infty$ to
$-\infty$), $\phim$, namely,
$\phic\equiv(\phip+\phim)/2$ and $\phi_{\Delta}\equiv \phip-\phim$,
respectively.  
$\phip$ and $\phim$ should be identified with each other in the end.
The kernels in (\ref{effectiveaction}) are defined by
\bea
C(x-x')  &\equiv&\frac{\lambda^2}{\mg^2}
\eta^{\mu_1\nu_1}
\eta^{\mu_2\nu_2} \im\lkk
\CG_{\mu_1\mu_2}^{F}(x-x')\CG_{\nu_1\nu_2}^{F}(x-x')\rkk  \:,~~{\rm for}
~t-t' > 0,
 \label{35} \\
D(x-x') &\equiv&\frac{\lambda^2}{2\mg^2}
\eta^{\mu_1\nu_1}
\eta^{\mu_2\nu_2} \re\lkk
\CG_{\mu_1\mu_2}^{F}(x-x')\CG_{\nu_1\nu_2}^{F}(x-x')\rkk .\non\\
\eea
Here $\CG_{\mu\nu}^{F}(x,x')$ is a finite-temperature
Feynman propagator of field derivatives defined by
\bea
\CG_{\mu\nu}^{F}(x,x')\equiv\langle \beta|T\partial_\mu\chi(x)
\partial_\nu\chi(x')|\beta\rangle
 =\partial_{x^\mu}\partial_{x'^\nu}G_\chi^{F}(x-x')+i\delta_{\mu
 0}\delta_{\nu 0}\delta^4(x-x'),
\eea
with $G_\chi^{F}(x)\equiv \langle\beta|T\chi(x)\chi(x')|\beta\rangle$.

The effective action (\ref{effectiveaction})
is complex-valued as a manifestation of
the dissipative nature of the system.  We cannot obtain any sensible
equation of motion by simply differentiating with respect to a field
variable because we are dealing with a real scalar field and its
equation of motion should be real-valued.  As shown in  \cite{Mor},  
one can obtain a real-valued effective action by introducing
a random Gaussian variable $\xi(x)$, which represents fluctuation
related to dissipation, with the dispersion 
$\la\,\xi(x)\xi(x')\,\ra = D(x-x')$.
As a result the equation of motion for the expectation value
$\phi_{c}(x)$ is given by
\bea
 (\,\Box+m_\phi^2\,)\,\phi_{c}(x)
          + \int_{-\infty}^{t}dt'\int d^3x'C(x-x')\phi_{c}(x') 
           =\xi(x) \:. \label{lineareq}
\eea
  Hereafter we omit the
suffix $c$.

As described in 
\cite{Yokoyama:2004pf,Yokoyama:2005dv}, this equation can easily be
solved using Fourier transform.  As a result, we find
the dissipation rate of zero-mode modulus oscillation is 
related to the imaginary part of $\phi'$s self energy and 
given 
by
\beq
  \Gamma_\phi=i\frac{\tilde C_0(m_\phi)}{2m_\phi}, \label{decayneww}
\eeq
with $\tilde C_0(m_\phi)$ being the $\vek=0$ mode of the
Fourier transform, $\tilde C_{\vecs k}(\omega)$,
of the memory kernel $C(x)$.  


In order to take thermal effects of $\chi$ correctly into account,
we should use the full dressed propagator,
$\CG_{\mu\nu}^{F({\rm drs})}(x,x')$, to calculate 
$\tilde C_{\vecs k=0}(\omega=m_\phi)$  \cite{Yokoyama:2005dv}.  
It is obtained by  calculating 
\beq
  \CG_{\mu\nu}^{F({\rm drs})}(x,x')=\langle \beta |T\partial_\mu\chi(x)
\partial_\nu\chi(x')\exp\lmk -i\int \CL_{int}d^4x\rmk|\beta\rangle,
\eeq
using the Matsubara representation \cite{Matsubara:1955ws} 
and resummation.
 As a result, the spatial Fourier mode of
the finite-temperature
dressed propagator
is given in terms of the spectral representation as
\beq
 \CG_{\mu\nu}^{F({\rm drs})}(\vep,t)=i\int \frac{d\omega}{2\pi}
\lmk\lnk\lkk 1+n_B(\omega)\rkk\theta(t)
+n_B(\omega)\theta(-t)\rnk\lkk
\frac{1}{(\omega+i\Gamma_p)^2-\omega_p'^{2}}
-\frac{1}{(\omega-i\Gamma_p)^2-\omega_p'^{2}}\rkk p_\mu p_\nu 
+i\delta_{\mu 0}\delta_{\nu 0}\rmk e^{-i\omega t}, \label{dress}
\eeq
with $p_\mu=(\omega,\vep)$,
$\omega_p'^{2}=\vep^2+m_\chi^2+\Sigma_R(p)+\Gamma_p^2$ and
$\Gamma_p=-\Sigma_I(p)/(2\omega)$, where $\Sigma_R(p)$ and $\Sigma_I(p)$
are real and imaginary parts of $\chi$'s self energy.
The real part typically scales as $\Sigma_R(p)\sim g^2T^2$.

Inserting (\ref{dress}) into (\ref{decayneww}),
the dissipation rate of the coherent field oscillation is given by
\begin{eqnarray}
\Gamma_\phi=&&\!\!\!
\frac{\lambda^2}{2m_\phi\mg^2}\int\frac{d^3p}{(2\pi)^3}
\frac{1}{4\op^2}\lnk (2\vep^2+m_T^2)^2[2n_B(\op)+1]
\lkk\frac{2\Gamma_p}{(m_\phi-2\op)^2+(2\Gamma_p)^2} -
\frac{2\Gamma_p}{(m_\phi+2\op)^2+(2\Gamma_p)^2}\rkk\right. \nonumber \\
&&+\lkk -2(2\vep^2+m_T^2)^2n_B^2(\op)e^{\beta\op}\beta\Gamma_p
+4(2\vep^2+m_T^2)\op\Gamma_p[2n_B(\op)+1]\rkk \nonumber \\
&&\times\lkk \frac{m_\phi-2\op}{(m_\phi-2\op)^2+(2\Gamma_p)^2}+
\frac{m_\phi+2\op}{(m_\phi+2\op)^2+(2\Gamma_p)^2}\rkk 
+\left.2m_T^4n_B^2(\op)e^{\beta\op}\beta\Gamma_p
\frac{2m_\phi}{m_\phi^2+(2\Gamma_p)^2}\rnk,  \label{22}
\end{eqnarray}
to the first order in $\Gamma_p$.  Here $m_T^2=m_\chi^2+\Sigma_R$ 
is the finite-temperature mass.
We find that the first two terms are 
 at most of order of $T^3/\mg^2$ for $\Gamma_p \lesssim T$.
The last term is also of the same order of magnitude 
if $\Gamma_p$ is much larger than $m_\phi$
 \cite{bo}, but 
it may have a different dependence and could
be larger for a sufficiently {\it small} $\Gamma_p$.  So we 
concentrate on it hereafter.

Assuming that there are $N$ degrees of decay modes with
the same strength, the last term of (\ref{22}) reads
\beq
  \Gamma_\phi = \frac{N\lambda^2}{2\mg^2m_\phi}
\int_{m_T}^{\infty}\frac{ d\omega}{2\pi^2\omega} 
\sqrt{\omega^2-m_T^2}m_T^4n_B^2(\omega)e^{\beta\omega}\beta
\frac{2m_\phi\Gamma_p}{m_{\phi}^2+(2\Gamma_p)^2}.
\label{dissipationrate}
\eeq
The last factor is maximal and equal to 1/2
 when $m_\phi = 2\Gamma_p$.  In this situation (\ref{dissipationrate})
is given by
\beq
  \Gamma_\phi \approx \frac{N\lambda^2 m_T^3T}{16\pi m_\phi \mg^2}
  \sim \frac{T^4}{\mg^2 m_\phi}.
\eeq
Hence it could be enhanced by a factor of $T/m_\phi$ compared with
other terms.  Indeed it can be larger than the Hubble parameter,
$H\approx T^2/\mg$, if 
\beq
  \frac{N\lambda^2 g^3}{16\pi} \gtrsim \frac{m_\phi}{H},
\eeq
where we have replaced $m_T$ by $gT$.
Since the right-hand-side is larger than unity in the oscillation
regime, from the above condition we need $N \gg 16\pi/(\lambda^2
g^3)$.  That is, assuming the coupling constants are of order 
unity, $N$ should be larger than 50 or so.  

Thus in such a 
rather special situation, thermal effects may dissipate the 
moduli oscillation efficiently.  For this to be the case
the imaginary part of the thermal correction to the mass of the
decay
product $\Gamma_p$ must be sufficiently small to satisfy
$\Gamma_p \sim m_\phi$.

In summary, we have calculated thermal effects on the dissipation
of modulus oscillation and have shown that they can dissipate
its energy efficiently in some special situations.

\acknowledgments{ 
This work was partially supported by the JSPS
  Grant-in-Aid for Scientific Research No.\ 16340076.}

\end{document}